\documentclass[epj,referee]{svjour}
\usepackage{graphicx}
\newcommand{\beq}{\begin{equation}}
\newcommand{\eeq}{\end{equation}}
\newcommand{\bea}{\begin{eqnarray*}}
\newcommand{\eea}{\end{eqnarray*}}
\newcommand{\bnea}{\begin{eqnarray}}
\newcommand{\enea}{\end{eqnarray}}
\newcommand{\p}{\partial}
\begin{document}
\title{Nonlocal feedback in nonlinear systems}
\author{R. Zambrini\inst{1} \and F. Papoff \inst{2}}

\institute{IFISC (CSIC-UIB), 
Campus Universitat Illes Balears, 07122 Palma de Mallorca, Spain \and 
SUPA,Department of Physics, University
of Strathclyde,  107 Rottenrow,  Glasgow G4 0NG, UK}
\date{Received: date / Revised version: date}
%
\abstract{
A shifted or misaligned feedback loop gives rise to a two-point nonlocality that is
the spatial analog of a
temporal delay. Important consequences of this nonlocal coupling have been found
both in diffusive and in diffractive
systems, and include convective instabilities, independent tuning of phase and group
velocities,
as well as amplification, chirping and even splitting of
localized perturbations.  Analytical
predictions about these nonlocal systems as well as their spatio-temporal dynamics
are discussed 
in one and two transverse dimensions and in presence of noise.
} 
\maketitle
\section{Introduction}
\label{intro}

There is a considerable interest in dynamical regimes in which small fluctuations
and "noise" are amplified.  In a large  class of optical systems this behavior
is caused by convective  instabilities~\cite{santagiustina97,taki00}. A
convective instability happens in presence of some source of drift when a state of
a nonlinear system becomes unstable and the group velocity of a localized
perturbation is larger than the velocity  of propagation of the instability
front. As a result, in the laboratory frame the perturbation is amplified but
moves away and eventually decays  at any point in a finite spatial domain if it
is not reflected at the boundary.  The existence of these instabilities has been
shown in systems in which the propagation of disturbances is characterized by
drift or walk-off, modelled by a gradient term. For such systems small regions of
convective instabilities  have been predicted and observed  in
hydrodynamics~\cite{ahlers83-gondret99},  
plasma~\cite{briggs} physics, traffic flow
~\cite{mitarai} and
optics~\cite{santagiustina97,louvergneaux04a,mussot08}. 

It was recently shown that  significantly larger
windows of convective instabilities are induced by nonlocal terms in the
governing equations~\cite{papoff05a}. In optics, these terms result from the
presence of an off-axis or shifted feedback loop which is modeled by a two point
nonlocality that is the spatial analogous of a temporally delayed feedback. This
is a common experimental issue and has been subject of both theoretical and
experimental study in liquid crystals light valves 
\cite{ramazza98a,rankin03a,pastur04a}, Kerr-like media 
\cite{seipenbusch97a,louvergneaux04a,agez06a} and
generic  nonlinear systems with diffusive \cite{papoff05a} and diffractive
\cite{zambrini07a} coupling. We note that the importance of feedback loops goes
well beyond the realm of optics \cite{arecchi99a} and has been long recognized in other fields of
physics and also in biology and engineering \cite{kitano01a,franklin02a}. Usually
feedback loops are introduced to better control the system and to limit the
growth of noise, while in the papers quoted a nonlocal feedback has been used
mainly to study fundamental properties of fluctuations in non linear systems when
convective
instabilities are induced. In
a recent experiment, however, an off-axis feedback loop has been used very
effectively to
suppress noise-sustained structures caused by an intrinsic drift term in a free
electron laser \cite{evain09a}, showing that two-point nonlocality are not only
fundamentally interesting, but also extremely useful.

\begin{figure}
\resizebox{0.75\columnwidth}{!}{\includegraphics{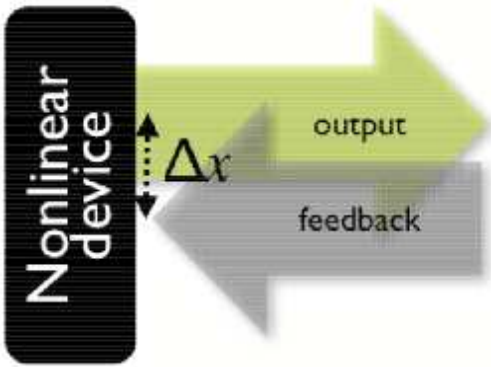}}
\caption{Schematic representation of a nonlinear device with shifted feedback.
The output beam is re-injected into the device after a shift $\Delta x$.}
\label{fig:0}       
\end{figure}

In this paper we review the effect of off-axis feedback in a broad class of
optical devices and explore the effect of the two-point nonlocality in the case of two
transverse dimensions. We consider passive as well as active nonlinear media with
fast decay of the polarization \cite{coullet89a}, including media with negative
refraction index  \cite{kockaert06a,hoyuelos} and devices with soft apertures
\cite{dunlop97b}.  Within the media we consider both diffusion and diffraction
and show how an off-axis feedback changes the nature of the first instability
threshold. As a consequence, there are large windows of control parameters
where  small localized signals can be strongly amplified while the background
radiation in other region of the system remains very low \cite{zambrini07a}.
The amplification does not need a continuous signal injection and takes place even
when the initial perturbation  is switched off.
Furthermore, in systems with diffraction and active media, the signal moves
across the  cavity with transverse phase and group velocities that are easily
managed to have the same or opposite signs \cite{zambrini07a}.  In spite of the
broken transverse reflection symmetry, localized perturbations can move both
towards or against the off-set direction and  can even split into two
counter-propagating components,  with the laser operating as a { signal
splitter}. Both noise sustained structures and signals control are  shown by
numerical simulations of the full nonlinear model confirming our theoretical
analysis and a rich spatio-temporal dynamics. Previous analysis
of Refs. \cite{zambrini07a,papoff09b}
are extended considering two transverse spatial dimensions.

\section{Equations}
\label{sec:1}

Off-axis feedback loops, in which the propagation of light is guided, are used with
nonlinear phase/amplitude modulators in which material evolves much more slowly than
the electric field and its polarization.  First examples were liquid crystal light
valves with feedback signal propagating through fibers whose position was easily
controlled \cite{ramazza98a,rankin03a,pastur04a}. These devices -in presence of a
shifted feedback- are modelled by equations of the type \cite{papoff05a,zambrini06a}
\bnea
\partial_{t}  \varphi  = \nabla^2_{\perp} \varphi + f_1(\varphi;\mu) + 
f_2(\varphi_{\Delta x};\mu) \label{dif_model}.
\enea
$\varphi(x,y,t)$ is a real field that represents the state of the material - for 
liquid crystal based devices it is related to the alignment of the molecules -  
at a point $(x,y)$ and time $t$, while $\varphi_{\Delta x}$ is evaluated  at  point $x+
\Delta x$ and time $t$, respectively scaled with
diffusion length and diffusion time. The control parameter $\mu$ is independent on $x$
for the sake of simplicity. $f_1$, $f_2$ are real functions that can
be derived with respect to  $\varphi$. In the limit of infinitely extended systems,
the homogeneous states  are  solutions of $f_1+ f_2=0$ and their domains of
existence depend upon $\mu$ but not upon $\Delta x$.

A similar feedback is used also with active materials inside optical cavities.
When the dynamics of  the difference of the populations of the energy levels
coupled to the light and of the polarization of the materials is much faster than
the dynamics of the electric field,  these systems are described by a single
equation of the type \cite{zambrini07a,papoff09b}
\beq
\p_t E = f(|E|^2;\mu)E + de^{i \delta} \nabla^2_{\perp} E + r e^{i \phi} 
E_{\Delta x}. \label{eq_classA}
\eeq
Here  $E$ is the slowly-varying amplitude of the electric field within a scalar
description,
$d\cos{\delta},d\sin{\delta}$  are the diffusion and diffraction
coefficients ($d>0$)and $\mu$ is a control parameter. Devices where only the 
polarization evolves
much faster than the electric field are instead modelled by  
\bnea
\p_t E &=& g_1(|E|^2,N; \mu)E + de^{i \delta} \nabla^2_{\perp} E + 
r e^{i \phi} E_{\Delta x} 
,\label{eqs_classB} \\ \nonumber
\p_t N &=& g_2(|E|^2,N,\nabla^2_{\perp}N;\mu), 
\enea
where $N$ represent the internal dynamics of the material coupled with light. We
consider feedback loops  in which the time delay $\Delta t$ is very small
compared to the time scale of the slowly varying envelope of the  field.  The
feedback can then be characterized by an amplitude $0<r<1$ and a phase shift 
$\phi = \omega_L \Delta t $, where $\omega_L$ is the carrier frequency and 
couples  the field $E$ in $(x,y)$ with the field $E_{\Delta x}$
in a shifted point.  $f$, $g_1$
and $g_2$ are nonlinear  complex functions that can be derived with  respect to
$E$ and $N$.  The trivial solutions  of Eq. (\ref{eq_classA}) and 
Eqs.(\ref{eqs_classB}) are $E=0$, and $E=0$, $N=N_0$, respectively. Interestingly, our
analysis applies also to  equations  with the more general feedback term
$[f_1(|E|^2)E]_{\Delta x}$, with  $r e^{i \phi}=f_1(0)$.

We are interested in the dynamics of  perturbations 
$\delta E \sim \exp{(\omega t +i {\bf k} \cdot {\bf x})}$ around the uniform states 
in the 
linear regime. The dynamics of perturbations for Eq.(\ref{eq_classA}) and 
Eqs.(\ref{eqs_classB}) is contained in the dispersion
relation 
\beq
\omega = \beta - e^{i \delta} (k_x^2+k_y^2) +r e^{i (\phi +k_x \Delta x)},
\label{ap_disp}
\eeq
where space is rescaled in units of $\sqrt{d}$. 
Here $\beta = f(0)$ for Eq.(\ref{eq_classA}) and 
$\beta = g_1(0,N_0)$  for  Eqs.(\ref{eqs_classB}). 
For class B models, perturbations $\delta N$ are always 
damped and decoupled from $\delta E$ and can be ignored.

We note that Eq.(\ref{ap_disp}) with 
$\delta= 0$,  
$\beta =
\p_{\varphi}f_1$, $r=|\p_{\varphi}f_2|$ and $\phi=0$ or $\pi$ depending on the sign of
$|\p_{\varphi}f_2|$, is the dispersion relation for the
perturbations of the uniform states of Eq.(\ref{dif_model}). This shows that
despite the  different physical meaning of the variables $\varphi$ and $E$, as well
as the significant differences in the characteristic time scales  and in the
light-matter coupling behind Eq.(\ref{dif_model}) and Eq.(\ref{eq_classA}) or
Eqs.(\ref{eqs_classB}), the dynamics of the perturbations of
Eq.(\ref{dif_model}) is a special case of the dynamics of the perturbations of
the other two cases. 

From Eq.(\ref{ap_disp}) we find that there are unstable band of ${\bf k}=(k_x,k_y)$
with 
the most unstable ones given by
\bnea
\nabla_{k} \omega_R & = &  -(2 k_x \cos{\delta}  + r \Delta x 
\sin{(k_x \Delta x + \phi)},2 k_y \cos{\delta} )  \nonumber \\
      & = & (0,0) \label{ki}\\
\p^2_{k_x^2} \omega_R & = &  -(2 \cos{\delta} + 
r \Delta x^2 \cos{(k_x \Delta x + \phi)}) <0, \label{sec_der}\\
\p^2_{k_y^2} \omega_R & = & -2 \cos{\delta} <0. \label{sec_der_y}
\enea 
The subscripts $R$ and $I$ refer to real and imaginary part, respectively.
The conditions (\ref{sec_der}-\ref{sec_der_y}) ensure that the solution of
Eq.(\ref{ki}) corresponds to the perturbation with the largest amplification; note
that  Eq.(\ref{sec_der_y}) is the standard stability condition for diffusive
equations.

\begin{figure}
\resizebox{0.75\columnwidth}{!}{\includegraphics{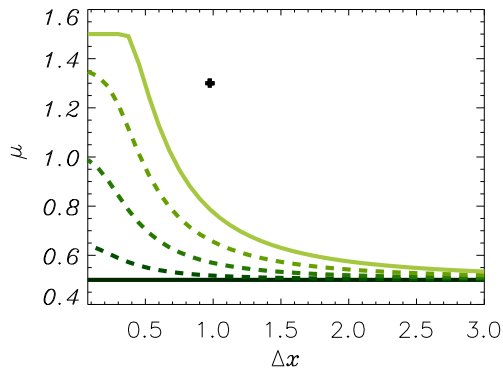}}
\caption{Instability threshold as a function of the shift.
The control parameter  $\mu=\beta_R+1$ is represented for easy comparison with results 
in the following section.
The effect of the feedback phase on the instability 
is shown: thresholds for $\phi=0$ (horizontal dark line) and increasing values
$\phi=\pi/4,\pi/2,3\pi/4,\pi$ (lighter colors), $\delta=0.49\pi$, $r=0.5$.
The dot corresponds to the parameter choice in Fig. \ref{fig:2_signals}.}
\label{fig:1}            
\end{figure}

An important feature of the instability threshold is that it is a function  of four
relevant parameters, namely $\phi, \delta, r\Delta x^2$, and  $\beta_R\Delta x^2$, and
is independent on $\beta_I$; 
therefore increasing the shift size  $\Delta x$ produces on the device the same effect
of larger gain $\beta_R$  and feedback $r$. A specific effect of the nonlocality is
that the relative strength of diffusion and diffraction, $\delta$, also becomes an
effective parameter to control the threshold position.  For active materials, the
lowest gain and feedback thresholds are generally found in the purely diffractive
limit ($\delta\sim\pi/2$).  The threshold value for  the scaled feedback strength
$r\Delta x^2$ is independent on the sign of the refractive index (sign of $\delta$)
and increases  with diffusion. Both $ \beta_R$ and $r$ can be increased to cross the
laser threshold and --similarly to the case of perfect alignment-- if the feedback is
out of phase then stronger gain is required,  as shown in Fig. \ref{fig:1}. For
vanishing shift ($\Delta x=0$) this phase acts as a detuning and increases the
threshold while for larger shift values the threshold tends asymptotically to the
value for in-phase feedback ($\phi=0$). Quite surprisingly, a misalignment lowers the
threshold because the  most  unstable mode has  $k_{x,I} \ne 0$ and, in this case, the
nonlocal coupling can   reduce the feedback dephasing. Consistently with this
interpretation, when the feedback perfectly in phase with the intracavity field
($\phi=0$) the most unstable mode is  homogeneous with $k_{x,I}=0$ and the threshold is
independent on the lateral shift $\Delta x$.  

\begin{figure}
\resizebox{0.75\columnwidth}{!}{\includegraphics{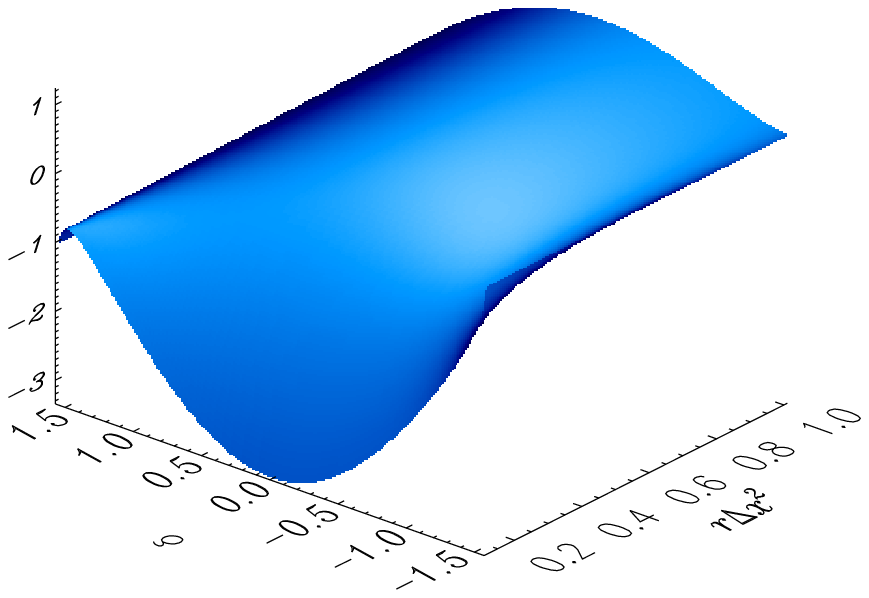}}
\resizebox{0.75\columnwidth}{!}{\includegraphics{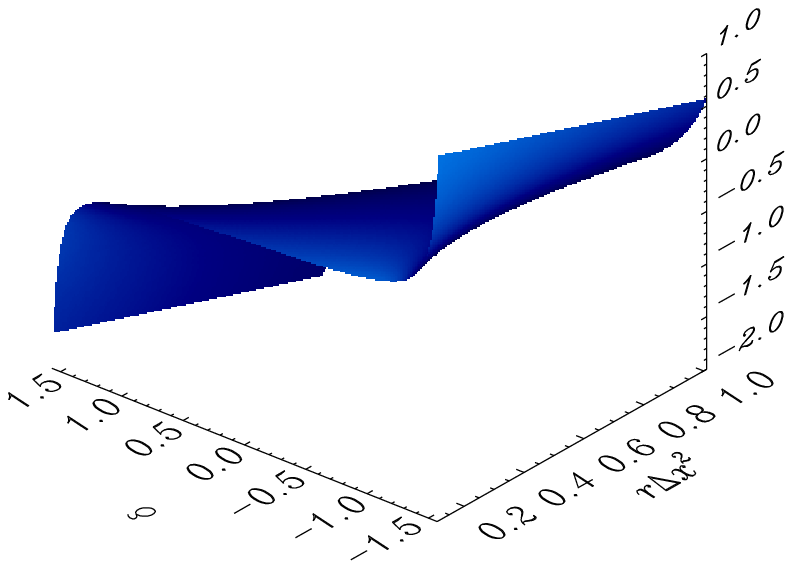}}
\caption{Variation of phase (upper plot) and group (lower plot) velocies, 
$v_{p,x}(k_c)$ and $v_{g,x}(k_c)$ respectively, depending on
 $\delta$ and $r\Delta x^2$. Other
parameters: $\beta=0.2-i0.02$ and $\phi=\pi/4$.}
\label{fig:3}            
\end{figure}

The dispersion Eq. (\ref{ap_disp}) has in general a not vanishing imaginary part
corresponding to  a non null phase velocity 
\bnea
{\bf v}_p & = & -\omega_I({\bf k}) \frac{{\bf k}}{|{\bf k}|^2}=
\\ \nonumber 
& - &[\beta_I -( k_x^2 +k_y^2) \sin{\delta} + 
r \sin{(k_x \Delta x +\phi)}]\frac{{\bf k}}{|{\bf k}|^2},
\enea
For the most unstable ${\bf k}$, ${\bf k}_c$, the phase velocity is
\beq
{\bf v}_p({\bf k}_c) =  ({k_x}_c \sin{\delta}   -  \frac{\beta_I}{{k_x}_c}  + 
\frac{2 \cos{\delta}}{\Delta x}, 0), \label{ph_vel}
\eeq
and group velocity is
\bnea
{\bf v}_g({\bf k}_c) & = & \nabla_k \omega_I({\bf k}_c)=  \nonumber \\
& &(2 {k_x}_c \sin{\delta}  
- r \Delta x \cos{({k_x}_c \Delta x + \phi)},0)=
\nonumber \\
& & (2 {k_x}_c \sin{\delta} \mp 
\sqrt{1-\frac{4 {k_x}_c^2 \cos{\delta}^2}{r^2 \Delta x^2}},0).
\label{v_g}
\enea
Note that both ${\bf v}_p({\bf k}_c)$ and ${\bf v}_g({\bf k}_c)$ are null in the direction orthogonal of the shift
and coincide with the one dimensional values.
In Eq. (\ref{v_g}) the $-$ sign solution always satisfies Eq.(\ref{sec_der}) and therefore
corresponds to
a wave-packet with frequency spectrum centered on a local maximum of the
amplification. On
the contrary, the $+$ sign may not satisfy Eq.(\ref{sec_der}). We remind that 
Eqs.(\ref{ph_vel}-\ref{v_g}) with  $\delta=0$, $\phi=0$ or $\phi=\pi$
give the velocities for the diffusive systems of 
Eq.(\ref{dif_model}).  An example of the large variability of phase and group
velocities at the critical wave-number are shown in Fig. \ref{fig:3}.        

The analysis above shows that the phase velocity at ${\bf k}_c$ and the group velocity are
always parallel to the shift, while their sign varies. There are manifolds in
the control parameter space that separate regions in which the group and the
phase velocity of the most unstable perturbation have the same  sign from region
in which these velocities have opposite sign.  Moreover, the real part of the
dispersion relation may have more than one maximum, so that a single
perturbation may split into two wave-packets. The independent tunability of phase
and group velocity is a specific feature of optical systems with two-point
nonlocality: as a matter of fact, gradient terms fix both velocities in the same 
direction of the drift, while two-point nonlocality in diffusive systems gives
always opposite velocities at the critical wavenumber ($v_p({\bf k_c})=- v_g({\bf
k_c}) $).
 
In the purely diffractive limit $\delta\rightarrow \pi/2$, both velocities are
actually odd functions of $k_c$; this symmetry is reduced by the effect of
diffusion ($|\delta|<\pi/2$). Therefore, even if for  $\phi=\pi$  both $+k_c$
and $-k_c$ are unstable, from the linear analysis we do not expect intensity
stripes above threshold in such optical systems.  As a matter of fact, the instability of
these two states (different traveling waves) is rather peculiar and opens the
possibility of   bistability instead of stripe pattern formation. 

Note that the tunability
of transverse phase and group velocities is a general
property that is valid also in the case in which $\Delta t$ is of the order of
the time scale of the slowly varying amplitude~\cite{papoff09b}.
This tunability is therefore a rather robust and distinctive feature of two-point
nonlocality with respect to models where velocities are induced by drift terms 
\cite{zambrini06a}. We remark that the parameters $\phi$ and $\delta$
are essential for this tunability: for this reason in the diffusive systems of 
Eq.(\ref{dif_model}) the tunability is absent and the phase and group
velocities have always opposite sign.
\begin{figure}
\resizebox{0.95\columnwidth}{!}{\includegraphics{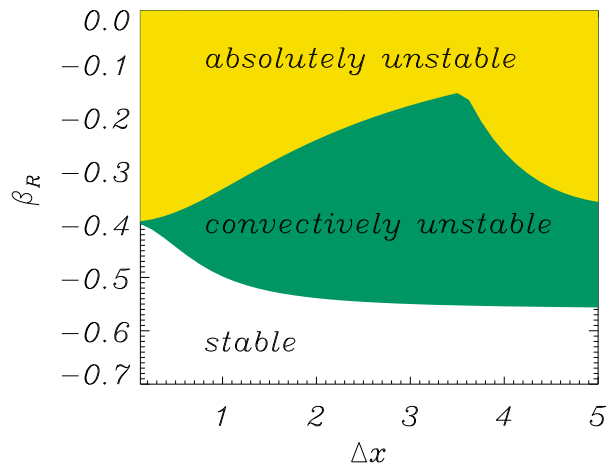}}
\caption{Regimes in which the homogeneous solution is
stable, convectively unstable and absolutely unstable as a function of the shift and 
for $\delta=0.45\pi$, $r=0.56$,
$\phi=0.25\pi$.}
\label{fig:2}            
\end{figure}

We now consider whether the perturbations with non null group velocities 
grow fast enough to occupy the entire system (absolute
instability) or if their group velocity is such that the perturbations, although
growing, move away (convective instability). 

The analysis is performed in an infinite system evaluating the asymptotic behavior
of perturbations both in a traveling and in a fixed reference frame.  Note that the
distinction between absolutely and convectively unstable regimes is generally given
in infinite system while the inclusion of boundary effect in finite systems can
change, in some case even drastically, the instability scenario. The simpler
example is the case of periodic boundary conditions: in this case no convective
instability does exist at all. Another interesting related question for finite
systems is   the phenomenon of transient growth of perturbations observed in cases
where the linear stability operator is non normal, i.e. does not  commute with its
adjoint \cite{papoff08a}. The connection between convective instability and   
transient growth of perturbations is still an open question \cite{chomaz}. 
What has been characterized is the  macroscopic amplification of
quantum noise in optical systems in this regime \cite{zambrini02}.

We determine absolute thresholds by evaluating asymptotically the Green function: 
we extend analytically the dispersion relation to  complex wavevectors $k$ (for
one transverse dimension) and  find
the appropriate integration paths in the plane $k$.  A detailed analysis is given in
\cite{papoff09b}, here we remark only that  the for the purely diffusive systems of
Eq.(\ref{dif_model}) the asymptotic evaluation of  the Green function is done by
closing the integration contour using only equiphase lines  from saddle points.
Furthermore, it turns out that the absolute threshold is determined only by the
saddle closest to the imaginary axis.
For Eq.(\ref{eq_classA}) and Eqs.(\ref{eqs_classB}) instead, we  close with steepest
descent paths a finite segment $[{k_I}_m,{k_I}_M]$ of the imaginary axis containing
all the $k_I$ with $w_R(0,k_I)>0$ and with ${k_I}_m$, ${k_I}_M$, such that
$w_R(0,k_I)<0$ for  $k_I \le {k_I}_m$ and $k_I \ge {k_I}_M$. For these two cases, the
correct  determination of the absolute thresholds requires to identify the
integration paths and to evaluate the contribution to the Green functions of the
saddle points that are part of it, excluding all the others.

Summarizing, our analysis allows to distinguish three regimes both in diffusive 
\cite{papoff05a} and laser \cite{zambrini07a} systems with a two-point nonlocality,
as shown
in Fig. \ref{fig:2} for a particular parameters choice.  The homogeneous (vanishing or
non-lasing) state becomes unstable above a first threshold (convectively unstable
regime)
corresponding to positive dispersion Eq.(\ref{ap_disp}). The absolutely unstable
regime is
found only after evaluation of the asymptotical growth of perturbations, involving an
integral whose approximate value is found with a non-trivial application of 
the saddle-point technique. This calculation
is  fully described in  \cite{zambrini07a}
and \cite{papoff09b} and here we only stress that not monotonic threshold
dependence on the shift can be found (Fig. \ref{fig:2}) and have  been also checked  by
numerical simulations of dynamical equations.

\section{Nonlinear and stochastic spatio-temporal dynamics}
\label{sec:2}

The general analysis of the previous section encompasses a broad class of
non-diffractive as well as laser models. Main features of the spatio-temporal
dynamics of specific systems can be anticipated from the linear stability analysis
and we will see some examples in the case of a class A 
\cite{jakobsen94a} laser:
\begin{eqnarray}\label{eq:laser}
\frac{\partial}{\partial t} E &=& - E (1+i\theta-N) + e^{i\delta} \nabla_\perp^2 E +
 r e^{i\phi}  E(x+\Delta x,y) +\nonumber \\
 &&\epsilon\xi(x,y,t) \nonumber\\
 N&=&\frac{\mu}{ 1+|E|^2}
\end{eqnarray}
with $\xi(x,y,t)$ complex Gaussian white noise. Numerical simulations of this
nonlinear model are
performed with a second-order in time Runge-Kutta method and using the random number
generator of
Ref.~\cite{Toral93}. The connection with the analysis of the previous section is
given by
$\beta=\mu - 1 - i \theta$. 
%
%
Numerical simulations  confirm  the predicted
instability diagram;
the wavenumbers dynamically selected and the velocities are well approximated by
those obtained
from the analytical analysis of the linear dispersion.

\subsection{Signal splitting and interactions}
\label{sect:signals}

\begin{figure}
\resizebox{0.75\columnwidth}{!}{\includegraphics{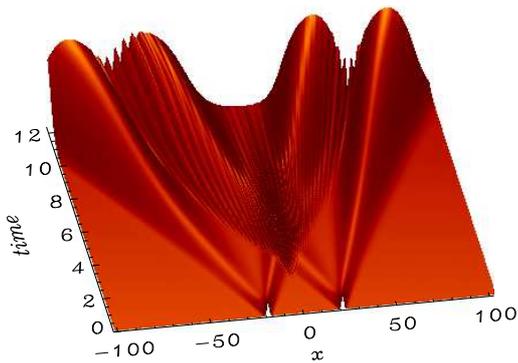}}
\caption{Spatio-temporal $intensity$ dynamics showing the interaction of signals in a 1D class A
laser with parameters $\Delta x=1$, $\mu=1.3$, $\theta=-1$, $\delta=0.49\pi$, 
$r=0.5$ , $\phi=\pi$, $\epsilon=0$ (see Fig. \ref{fig:1}). 
Evolution starting from two small Gaussian perturbations
as can be recognized in the picture  at time $t=0$.}
\label{fig:2_signals}            
\end{figure}
The regime of convective instability allows the amplification of localized light
signals below the lasing threshold. The direction of propagation of these signals
depends on the group velocity Eq. (\ref{v_g}) 
and steering can be obtained also non-mechanically
by varying the feedback loop phase, as shown in Figs. 3 and 4 of Ref.
\cite{zambrini07a}.
Particularly interesting is the case in which the feedback is negative, $\phi= \pi$,
first discussed in Ref. \cite{zambrini07a}. The recovered symmetry in the dispersion
relations gives rise to instability of both positive and negative wave-vectors and,
as a peculiar consequence of the two-point nonlocality, these waves propagate in
opposite directions. This allows the system to operate as a signal splitter in which
an initial perturbation, such as a localized spot of light, is divided into two
counter-propagating copies. This phenomenon is robust also in presence of noise
 \cite{papoff09b} and deviations from the symmetry, i.e. differences between
 the left and the right propagating signals, are due to nonlinear effects
 \cite{zambrini02,zambrini05}.

Once excited, these localized signals are amplified and propagate 
in the system (even in opposite directions) without giving rise to any 
intensity 
modulation, even if the field is actually spatially oscillating. 
We will now focus on a different aspect of signal
control, namely signal interaction when they cross each-other. 
As localized perturbations are split into
counter-propagating ones it is actually possible to have signals crossing during
temporal evolution. This case is represented in Fig. \ref{fig:2_signals}. In
this example  the system is excited in two separated points at an initial time
and the dynamics is considered in the case of only one transverse  dimension. We
see that two of the four generated signals cross and locally interfere
generating a transient intensity stripe in the center of Fig.
\ref{fig:2_signals}. Moreover, looking on the left (right) side, we see that 
due to the different velocities of perturbations fronts, the trailing edge of
the left signal is reached by the leading edge of the following signal and this
also generates intensity modulation by interference. Finally, after a transient
time (not shown in  Fig. \ref{fig:2_signals}) and as expected, due to the
convective character of the instability, the system evolves locally back to the
homogeneous vanishing state.

\subsection{Patterns in one and two dimensions}
 \label{sect:patterns}

\begin{figure}
\resizebox{0.8\columnwidth}{!}{\includegraphics{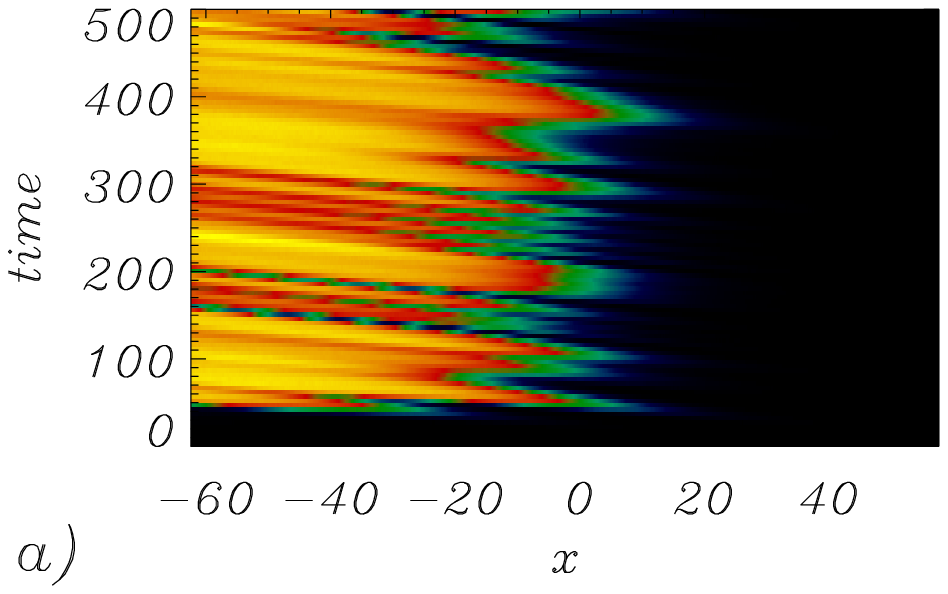}}
\resizebox{0.8\columnwidth}{!}{\includegraphics{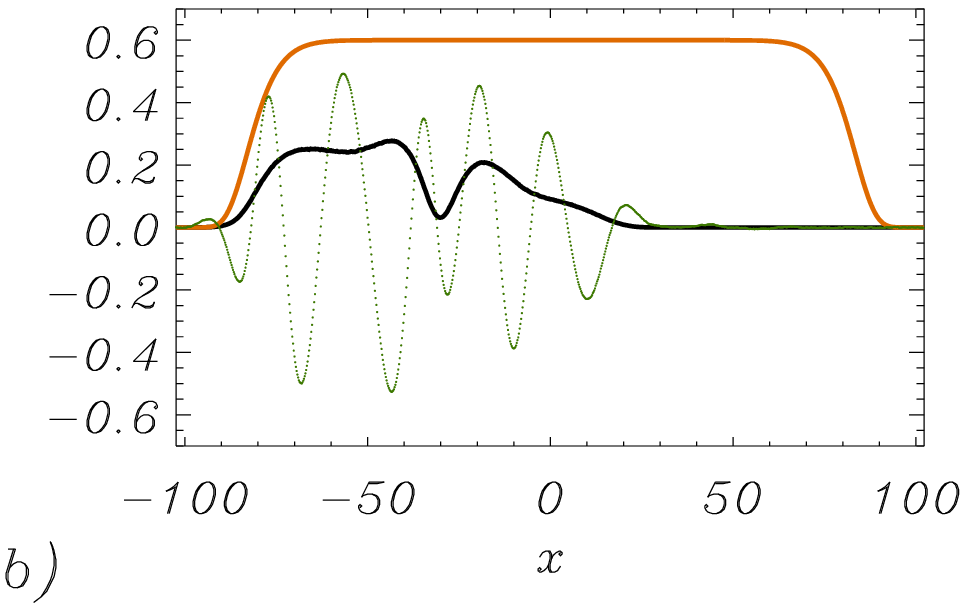}}
\caption{(a) Spatio-temporal evolution of the intensity field for one transverse
dimension and starting from a noisy initial condition. (b) Intensity $|E(x,t=100)|^2$
(black line), real part (green dotted line) of the field, and pump profile $\mu$
(orange line).  Parameters: $\Delta x=2$, $\theta=0$, $\delta=0.45\pi$,  $r=0.56$ ,
$\phi=\pi/4$, $\epsilon=10^{-3}$. Super-Gaussian profile of $\mu$ with maximum value 
$0.6$.}
\label{fig:1d}            
\end{figure}

In previous works we described the spatio-temporal dynamics of class  A lasers
assuming a one dimensional transverse geometry \cite{zambrini07a,papoff09b}. Good
agreement with the theoretical predictions (threshold position,  pattern wave-length,
phase and group velocity) was found when considering rather $large$ systems where
boundary conditions effects are negligible. The importance of the boundaries is
particularly evident in the convectively unstable regime. In Fig.~\ref{fig:1d} we show
the evolution of a noise sustained structure for $\delta=0.45\pi$, $r=0.56$,
$\phi=0.25\pi$ and super-Gaussian pump with maximum value $\mu=0.6$ in the central
region (see Fig.~\ref{fig:1d}b).  From Fig. \ref{fig:2}  and remembering that
$\beta_R=\mu - 1$, we see that for these parameters the nonlasing state is
convectively unstable and the field in Fig.~\ref{fig:1d}a) has the typical incoherent
profile discussed in  \cite{papoff09b}.  Here we show that a large system is needed in
order to observe noise sustained structures  intensities in the system:  the intensity
represented in Fig.~\ref{fig:1d}b is significantly high only in half of the system,
far from the right edge of the pump profile $\mu$. In other words, the intensity
growth --in this case against the shift direction-- is rather slow and large systems are
needed to observe an intense noise sustained pattern.
Oscillations appear in the field profile (phase pattern given by the green line in
Fig.~\ref{fig:1d}b).

\begin{figure}
\resizebox{0.95\columnwidth}{!}{\includegraphics{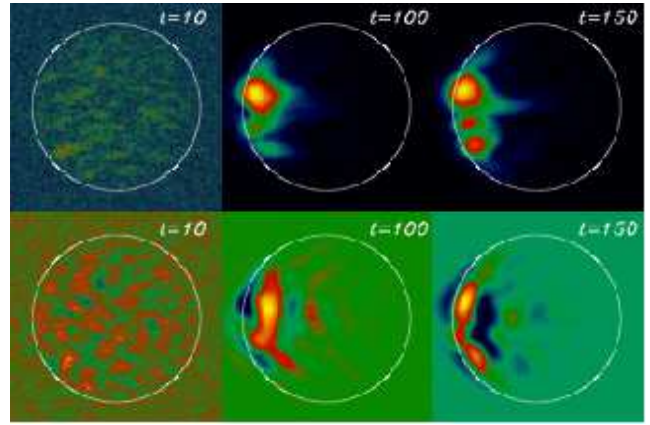}}
\caption{Temporal evolution of the intensity (upper line) and of the real part (lower
line) of the field starting from a noisy initial condition. Parameters: $\Delta x=2$,
$\theta=0$, $\delta=0.45\pi$,  $r=0.56$ , $\phi=\pi/4$, $\epsilon=10^{-3}$.
Super-Gaussian profile of $\mu$ with maximum value  $0.6$ inside the circular area. For
this value of $\mu$ the vanishing state is  convectively unstable.}
\label{fig:5}            
\end{figure}

In order to give a representation of the aspect of spatial structures as they would
appear in experiments with broad area lasers, it is interesting to consider the
evolution of two-dimensional field $E(x,y,t)$.  In Fig \ref{fig:5} we represent both
fields profiles and intensities for the same parameters as in Fig.~\ref{fig:1d},
showing the aspect of a noise sustained structure in the convective regime.
Three snapshots give an example of the dynamic character of this structure:
numerical simulations in presence of noise show an incoherent traveling structure
whose aspect is continuously changing.
\begin{figure}
\resizebox{0.95\columnwidth}{!}{\includegraphics{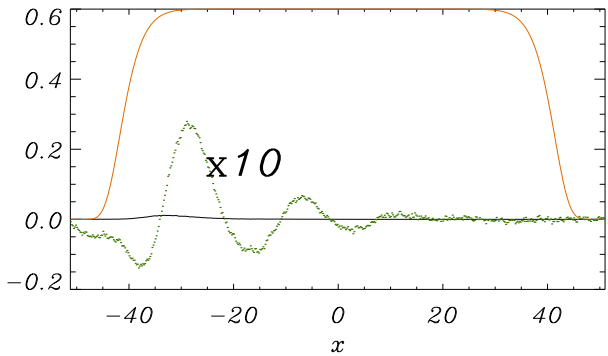}}
\caption{Section plot of the 2D intensity (black line) and real part (green dotted
line) of the field for $t=100$. The corresponding values are multiplied by a factor 10
to be compared with the pump profile $\mu$  (orange line). Same parameters of Fig.
\ref{fig:5}.}
\label{fig:5bis}            
\end{figure}
The importance of the boundary and system size
in this case is evident as it leads to a very low
intensity noise sustained structure. Due to the reduced size of the pumped area
with respect to previous 1D numerical simulations (compare $\mu$ super-Gaussians
in Figs.
\ref{fig:1d}  and  \ref{fig:5bis}) the intensity profile is dominated by a
smooth front emerging on the right side and remains very small. 

\begin{figure}
\resizebox{0.95\columnwidth}{!}{\includegraphics{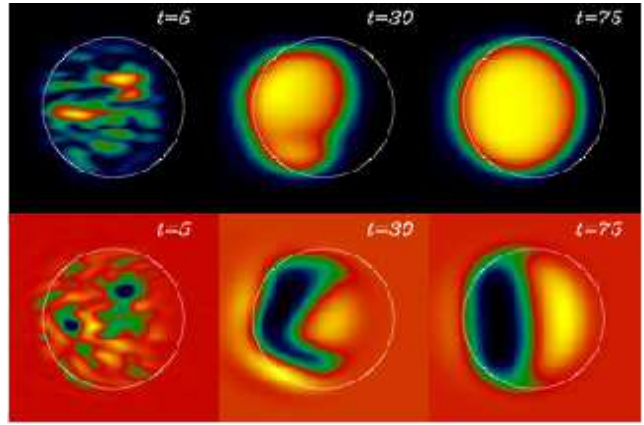}}
\caption{Temporal evolution of the intensity (upper line)
and of the real part (lower line) of the field for vanishing state absolutely unstable.
Parameters: $\Delta x=2$, $\theta=0$, $\delta=0.45\pi$, 
$r=0.56$ , $\phi=\pi/4$, $\epsilon=10^{-5}$. Super-Gaussian profile of $\mu$ with
maximum
value  $1$.}
\label{fig:abs}            
\end{figure}

When the pump is increased this front becomes steeper and and a uniform intensity state
occupies a broad
part of the system. Above the absolutely unstable threshold a coherent phase pattern
arises as shown in  Fig. \ref{fig:abs} and the system is in the lasing state.
Starting from a noisy initial condition, perturbations are washed away from the system
and  the intensity profile becomes intense and uniform. Also the coherence of the
underlying phase pattern increases (in this case the pattern aspect ratio is very
small and only one oscillation is actually seen in Fig. \ref{fig:abs}). Note also that
the intensity profile is deformed and displaced against the direction of the shift,
i.e. it is not exactly centered in the pumped region (white circle in Fig.
\ref{fig:abs}, panels at $t=75$).

\begin{figure}
\resizebox{0.9\columnwidth}{!}{%
  \includegraphics{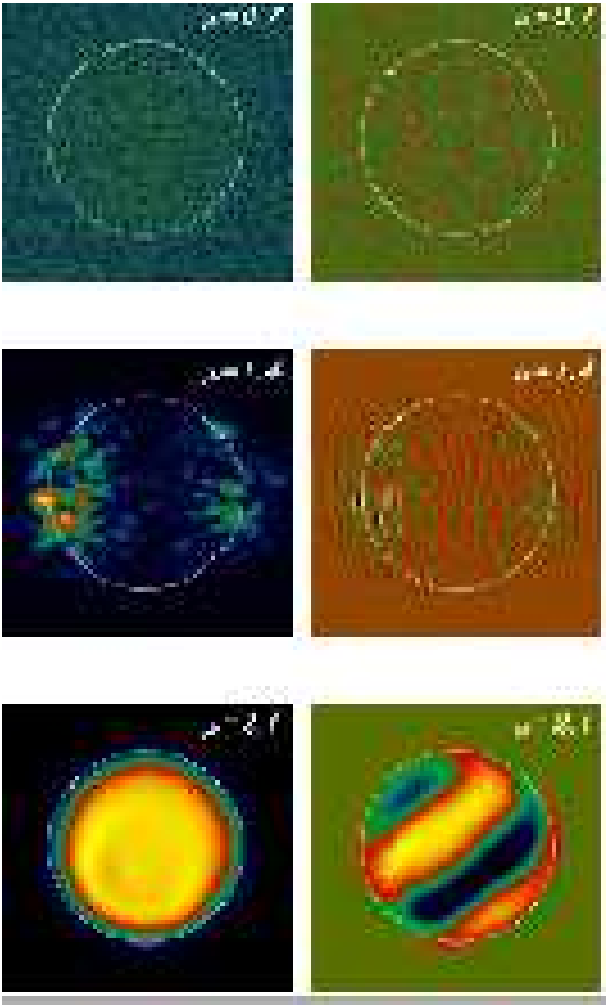}}
\caption{Spatial distributions of the intensity (left panels) and field real part
(right panels) for numerical simulations with super-Gaussian pump profiles with
maximum intensity $\mu=0.7,1.4$, and $2.1$ inside the circular panels. In the external
region the gain vanishes. Other parameters are
$\Delta x=1$, $\theta=0.2$, $\delta=0.49\pi$, 
$r=0.5$ , $\phi=\pi$, $\epsilon=10^{-5}$. Simulations are started from noise initial
condition.} 
\label{fig:4}            
\end{figure}

The most peculiar patterns in lasers with nonlocal feedback occur for  $\phi= \pi$,
when the system operates as a signal splitter. The change in the pattern aspect
when increasing the pump is shown in Fig. \ref{fig:4}. In the convectively unstable
regime (two panels with $\mu=1.4$) there are now two regions of larger intensity, as perturbations
with opposite wave-numbers travel apart, as expected. Intense light spots
correspond to a certain coherence in the underlying phase pattern, while the
intensity drops down where there are defects in the phase stripes. Intensity
reaches larger values on one side (here the left one, against the positive shift
direction,
$\Delta x=2$) 
\cite{zambrini07a,papoff09b,zambrini05}.

Noise sustained
patterns for negative feedback are then characterized by off-axis spots
and have a vanishing intensity in the central area, while
crossing the absolute instability threshold an intense and uniform profile is
reached -after a transient- in  the whole pumped region. Note that the phase pattern has
larger wave-numbers in the noise sustained structure. After a longer transient (not shown
in the picture) the
stripe in the absolutely unstable regime (Fig. \ref{fig:4} for $\mu=2.1$) becomes
orthogonal to the shift direction, as theoretically predicted.

\subsection{Conclusions}

Two-point nonlocality can be easily induced in optical devices through a displaced
feedback loop and leads to distinctive effects opening interesting possibilities in
experiments in extended devices. We have presented its analytical characterization
through linear approximations, allowing to predict instability convective and absolute
thresholds, unstable wave-numbers and velocity of drifting packets, in a broad class
of two-dimensional nonlinear systems. Many differences with respect to the
most studied drift, modelled by a gradient term, make this two-point nonlocality an
interesting and versatile tool in view of experiments. In
particular, such nonlocality opens significantly larger windows of control parameters
where the system output
state is sustained by the presence of noise instead of
the dynamics (convective regime). In such regime the system can be locally excited
generating a signal that is amplified during propagation. Note that the
amplification here considered occurs once the (initial) local perturbation is removed,
at difference of laser amplification above transparency, generally considered in
presence of a continuous signal injection.  

Another distinctive feature of displaced feedback and two-point nonlocality is the
possibility to independently tune phase and group velocity, while gradient terms fix
both velocities in the same  direction of the drift. In optical systems such
velocities have been shown to be either parallel or opposite and can be tuned even
non-mechanically through the feedback phase $\phi$ (without touching  the experiment
alignment). Such large tunability is a characteristic of optical systems, while
two-point nonlocality in diffusive systems gives always opposite velocities at the
critical wavenumber ($v_p({\bf k_c})=- v_g({\bf k_c}) $). Another distinctive feature
of two-point nonlocality in optical systems is the possibility to control the feedback
phase to operate the laser as a signal splitter (for $\phi=\pi$). Then an initial
perturbation is split in two copies amplifying (even when the injection is removed)
during their propagation in opposite directions through the broad area device. 

In general, no oscillations are observed in the intensity profile and only phase
patterns are present in lasers here considered. The interaction of traveling signals, however,
displays also intensity modulation in the interaction region, as discussed in Sect.
\ref{sect:signals}. Similarly, if instead of initial localized perturbations the system
is considered in presence of noise, a central area of intensity stripes can be
observed in a laser as shown in Fig. 10 of  Ref. \cite{papoff09b}. Numerical
simulations considering only one transverse dimension allow to easily visualize the
spatio-temporal dynamics, as in the case of the signals interactions in Fig.
\ref{fig:2_signals}, or to see the spatio-temporal coherence of  noise sustained
structures, as in  Fig. \ref{fig:1d}. On the other hand,  simulations in two transverse
dimensions here presented are important in view of  experimental realizations,
showing  pattern  features in the shift direction and in the orthogonal one. Examples
are the orientation of phase oscillations, the elongated aspect of transients domains,
as well as the displacement of lasing state with respect to the pumped region, as shown in
Sect. \ref{sect:patterns}. A last important aspect to be considered in view of
experiments is the system size: analytical predictions for infinite systems  are in
quantitative agreement with simulations for relatively large systems, while a finite
system effects gives important deviations in small systems not only
when few oscillations of
the phase patterns are present but also in the cases in which smooth fronts modulate
the light intensity profile.

Funding from FISICOS (FIS2007-60327) and  CoQuSys (200450E566) is acknowledged.


\begin{thebibliography}{}

\bibitem{santagiustina97} M.~Santagiustina, P.~Colet, M.~San Miguel, D.~Walgraef,
 Phys. Rev. Lett. {\bf 79} 3633 (1997); Phys. Rev. E {\bf 58}, 3843 (1998);
 Opt. Lett. {\bf 23}, 1167 (1998).

\bibitem{taki00} M.~Taki, N.~Ouarzazi, H.~Ward, P.~Glorieux,
JOSA B {\bf 17}, 997 (2000).


\bibitem{ahlers83-gondret99} 
J~ M.~Chomaz, Phys. Rev. Lett. {\bf 69} (1992) 1931;
K. L. Babcock, G. Ahlers, and D. S. Cannell, 
Phys. Rev. Lett. {\bf 67}  (1991) 3388; 
Instability
P.~Gondret, P.~Ern, L.~Meignin, M.~Rabaud, 
Phys. Rev. Lett. {\bf 82} (1999) 1442.


\bibitem{briggs} R. J. Briggs, {\it Electron-Stream Interaction with Plasmas} 
(MIT, Cambridge, MA, 1964)

\bibitem{mitarai} N. Mitarai and H. Nakanishi, 
\newblock Phys. Rev. Lett. {\bf 85}, 1766 (2000).

\bibitem{louvergneaux04a}
E.~Louvergneaux, C.~Szwaj, G.~Agez, P.~Glorieux, and M.~Taki,
\newblock Phys. Rev. Lett. {\bf 93} (2004) 101801.

\bibitem{mussot08} A. Mussot, E. Louvergneaux, N. Akhmediev, F. IT, Boston,
Reynaud, L. Delage, and M.~Taki,
\newblock Phys. Rev. Lett. {\bf 101} (2008) 113904.

\bibitem{papoff05a}
F.~Papoff and R.~Zambrini,
\newblock Phys. Rev. Lett. {\bf 94} (2005) 243903.

\bibitem{ramazza98a}
P.~Ramazza, S.~Ducci, and F.~Arecchi,
\newblock Phys. Rev. Lett. {\bf 81} (1998) 4128 .

\bibitem{rankin03a}
S.~Rankin, E.~Yao, and F.~Papoff,
\newblock Phys. Rev. A {\bf 68}  (2003) 013821.

\bibitem{pastur04a}
L.~Pastur, U.~Bortolozzo, and P.~Ramazza,
\newblock Phys. Rev. E {\bf 69}  (2004) 016210.

\bibitem{seipenbusch97a}
J.~P. Seipenbusch, T.~Ackemann, B.~B. B.~Sch\"apers, and W.~Lange,
\newblock Phys Rev. A {\bf 56}  (1997) R4401.

\bibitem{agez06a}
G.~Agez, P.~Glorieux, M.~Taki, and E.~Louvergneaux,
\newblock Phys. Rev. A {\bf 74} (2006) 043814.

\bibitem{zambrini07a}
R.~Zambrini and F.~Papoff,
\newblock Phys. Rev. Lett. {\bf 99} (2007) 063907.


\bibitem{arecchi99a}
F.~Arecchi, S.~Boccaletti, and P.~Ramazza,
\newblock Physics Reports {\bf 318} (1999)  1.


\bibitem{kitano01a} 
{\em Foundations of Systems Biology}, edited by H.~Kitano, MIT, Boston,  2001.

\bibitem{franklin02a}
G.~F. Franklin  et~al.,
\newblock {\em Feedback Control of Dynamic Systems},
\newblock Prentice Hall, 2002.

\bibitem{evain09a}
C. Evain, C. Szwaj, S. Bielawski, 
M. Hosaka, A. Mochihashi, M. Katoh, and M.-E. Couprie, 
Phys. Rev. Lett. \textbf{102}, (2009) 134501

\bibitem{coullet89a}
P.~Coullet, L.~Gill, and F.~Rocca,
\newblock Opt. Comm. {\bf 73}  (1989) 403.

\bibitem{kockaert06a}
P.~Kockaert, P.~Tassin, G.~V. der Sande, I.~Veretennicoff, and M.~Tlidi,
\newblock Phys. Rev. A {\bf 74}  (2006) 033822.

\bibitem{hoyuelos} D. A. Martin, M. Hoyuelos,
Phys. Rev. E  {\bf 80} (2009), 056601.

\bibitem{dunlop97b}
A.~Dunlop, W.~Firth, and E.~Wright,
\newblock Opt. Commun. {\bf 138}  (1997) 211.

\bibitem{papoff09b}
F. Papoff and R. Zambrini, Phys. Rev. A \textbf{79} (2009) 033811


\bibitem{zambrini06a}
R.~Zambrini and F.~Papoff,
\newblock Phys. Rev. E. {\bf 73}  (2006)  016611.


\bibitem{papoff08a}
F.~Papoff, G.~D'Alessandro, and G.-L. Oppo,
\newblock Phys. Rev. Lett. {\bf 100} (2008) 123905; G.~D'Alessandro, C.B.~Laforet, 
Opt. Lett.   \textbf{34}  (2009) 614; G.~D'Alessandro,F.~Papoff, Phys. Rev. A
\textbf{80} (2009) 023804.

\bibitem{chomaz} 
C. Cossu and J. M. Chomaz, Phys. Rev. Lett. {\bf 78} (2007) 4387.

\bibitem{zambrini02}
R.~Zambrini and M.~S. Miguel,
\newblock Phys. Rev. A {\bf 66}  (2002) 023807,
R.~Zambrini, S.~Barnett, P.~Colet, and M.~S. Miguel,
\newblock Phys. Rev. A {\bf 65}  (2002) 023813.

\bibitem{zambrini05}
R.~Zambrini, M.~S. Miguel, C.~Durniak, and M.~Taki,
\newblock Phys. Rev. E {\bf 72}  (2005) 025603(R).

\bibitem{jakobsen94a}
P.~K. Jakobsen, J.~Lega, Q.~Feng, M.~Staley, J.~V. Moloney, and A.~C. Newell,
\newblock Phys. Rev. A {\bf 49}  (1994) 4189.

\bibitem{Toral93}
R.~Toral and A.~Chakrabarti,
\newblock Comput. Phys. Commun. {\bf 74} (1993) 327.

\end{thebibliography}
\end{document}